# Medical Imaging and Machine Learning

Rohan Shad[1], John P. Cunningham[2], Euan A. Ashley[3,4], Curtis P. Langlotz[4,5], **William Hiesinger**[1,4]

[1]Department of Cardiothoracic Surgery, Stanford University; [2]Department of Statistics, Columbia University; [3]Department of Cardiovascular Medicine, Genetics, and Biomedical Data Science, Stanford University; [4]Stanford Artificial Intelligence in Medicine Center; [5]Department of Radiology and Biomedical Informatics, Stanford University

Abstract:

Advances in computing power, deep learning architectures, and expert labelled datasets have spurred the development of medical imaging artificial intelligence systems that rival clinical experts in a variety of scenarios.[1–8] The National Institutes of Health in 2018 identified key focus areas for the future of artificial intelligence in medical imaging, creating a foundational roadmap for research in image acquisition, algorithms, data standardization, and translatable clinical decision support systems.[9] Among the key issues raised in the report: data availability, need for novel computing architectures, and explainable AI algorithms, are still relevant despite the tremendous progress made over the past few years alone. Furthermore, translational goals of data sharing, validation of performance for regulatory approval, generalizability and mitigation of unintended bias must be accounted for early in the development process.[10] In this perspective paper we explore challenges unique to high dimensional clinical imaging data, in addition to highlighting some of the technical and ethical considerations in developing high-dimensional, multi-modality, machine learning systems for clinical decision support.

Introduction:

It is remarkably challenging to deploy AI systems that assist with even simple clinical tasks.[6,8] When taken out of siloed and controlled laboratory environments, users of AI systems must contend with input quality control, and devise ways to integrate these systems within established clinical workflows. Machine learning algorithms that were designed to *reduce* the time it took for clinically actionable inferences, when deployed in clinics resulted in patients inadvertently experiencing event greater delays.[11] These early forays into translatable clinical ML have shown that proactively identifying problems that may affect deployment within clinical workflows must begin at the inception of the design process. Extensive open source machine learning software libraries and advances in compute performance have made it easier for researchers to develop increasingly complex AI systems tailored towards specific clinical problems.[12,13] In addition to moving beyond detecting findings diagnostic for disease, the next generation of AI systems must account for systemic biases in training data, intuitively alert end-users to the uncertainty inherent in predictions, and allow for opportunities to explore and explain the mechanisms by which predictions are made. This perspective paper builds on these key priority areas for the acceleration of foundational AI research in medicine, and provides a template for researchers interested in navigating some of the issues and challenges that come with building clinically translatable AI systems.

**High Dimensional Medical Imaging Data**

We anticipate that the availability of high quality annotated medical datasets will continue to lag demand for the foreseeable future. Retrospectively assigning clinical ground truth labels require extensive investment of time from clinical experts, and there are significant barriers to aggregating multi-institutional data for public release.[9] In addition to 'diagnostic AI' characterized by models trained on hard radiological ground truth labels, there will be demand for 'disease prediction AI' trained on potentially more noisy clinical composite outcome targets.[8,14–16] Prospective data collection with standardized protocols for image acquisition and adjudication of clinical ground truth are essential steps towards building massive multicenter imaging datasets with paired clinical outcomes.

Working with large multicenter datasets brings with them a multitude of privacy and liability concerns associated with potentially sensitive data embedded in the imaging files themselves. The DICOM (Digital Imaging and Communication in Medicine) standard was designed to capture, store and provide workflow management for medical images, and is nearly universally adopted.[17] Imaging files (stored either as .dcm files or within DICOM_DIR) contain both raw pixel data and associated metadata. A multitude of open-source and proprietary tools exist to assist with de-identification of DICOM files.[9,18] Of note, backend frameworks such as the Google Healthcare API supports DICOM de-identification via 'safe lists'. On the user facing side, The MIRC Clinical Trials Processor (CTP) anonymizer is a popular alternative though it requires working outside of the python environment with legacy libraries.[18] The pydicom python package is both convenient and extensively documented, and can be used to process DICOM files prior to use or transfer to a collaborating institution.[19] Imaging data can then be extracted and stored as Numpy arrays, and relevant metadata stored on a separate csv file. While the arrays themselves can be stored as lossless pickle files, often it is more convenient and practical to store the imaging data as compressed .mp4 or .avi files.[20]

An oft-cited drawback with automated de-identification methods or scripts is the potential for 'burned in' PHI to remain on the imaging files. Despite the DICOM standard, manufacturer specific differences make it difficult to generate simple rules via tools such as MIRC CTP to mask out regions where PHI may be located. We suggest using a simple machine learning system for masking 'burned in' PHI. In the case of echocardiograms for example, there exists a pre-defined scanning sector where the heart is visualized. Other potential options are ML-based optical character recognition tools to identify and mask out regions with printed text. The DICOM tags themselves can be useful to extract both scan level information and modality specific tags. In the case of echocardiography and Cardiac MRI for example, important scan-level information such as acquisition frame rates and date, or MRI sequence (T1/T2) can readily be extracted from the DICOM metadata. Normalization of data is a critical pre-processing step in AI pipelines and normalizing framerates is often required in addition to standard pixel intensity normalization.

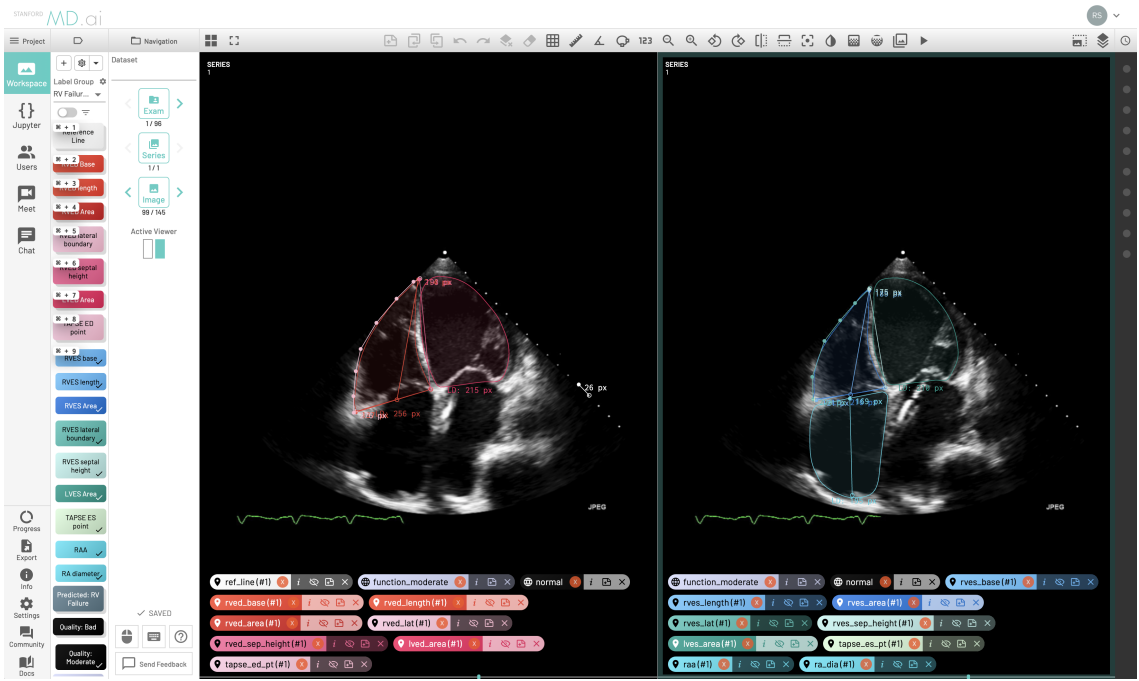

Figure 1. Cloud based tools such as MD.ai can be used for generating expert annotated datasets and evaluating against clinical experts via a secure connection.

For research endeavors that involve head-to-head benchmarking of AI systems against clinicians in an effort to establish equivalence or superiority, we recommend storing the scans in the DICOM format. This allows for deployment over scalable and easy to use cloud-based annotation tools. A number of solutions exist for assigning scans for assessment by clinical experts. The requirements may range from simple scan-level labels to detailed domain specific anatomical segmentation masks. At our institution, we deployed MD.ai (New York, New York) – a cloud-based annotation system that natively works with DICOM files stored on institutionally approved cloud storage providers (Google Cloud Storage or Amazon AWS). An additional advantage of this approach is that the scans are kept at native resolution and quality. Live real-time collaboration simulates 'team-based' clinical decision-making. Annotations and labels can easily be exported as .json files for downstream analyses. Most importantly, many of these tools are accessible remotely from any modern web browser, drastically improving user experience and reducing the technical burden on clinical collaborators.

Newer ML training paradigms such as federated learning may help circumvent many of the barriers associated with data sharing. Though there have been recent efforts to use federated learning in medical imaging, there remain significant technical challenges in implementing these methods.[21] Kaissis *et al.* review the principles, security risks, and implementation challenges with federated learning.[22] The key feature of this method is that local copies of algorithms are trained at each institution. At pre-determined intervals, the learned weights across each institutional algorithm are then pooled together and redistributed – effectively learning from a large multicenter dataset without the need to transmit or share any of the medical imaging data.[23,24] This has been instrumental in rapidly training algorithms to detect features of COVID-19 from CT-scans of the chest.[21]

**Computational Architectures**

Neural network architectures used in modern clinical ML are largely derived from those optimized for large photo or video recognition tasks.[25] These architectures have shown to be remarkably robust even in the otherwise challenging task of fine-grained classification, where classes have subtle intra-class variance (breeds of dogs) rather than obviously different objects with high inter-class variance (airplanes vs dogs).With adequate pre-training on large datasets (e.g. ImageNet) these 'off the shelf' architectures outperform their tailor made fine grained classifier counterparts.[26] Many of these architectures are available for use in popular machine learning frameworks such as TensorFlow and Pytorch.[27–31] The pytorch torchvision package for example, contains a library of image recognition neural network architectures. Most importantly, the package also provides ImageNet pre-trained weights for the networks, allowing researchers to rapidly repurpose these for specialized medical imaging tasks.[32]

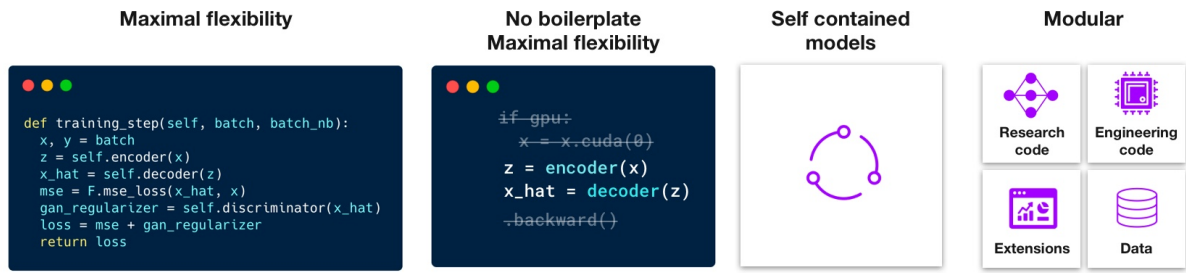

Figure 2: Modular approach to structuring machine learning code as described by the PyTorch Lightning team. Such frameworks enable rapid iteration and experiment management under a standardized framework, and bridges engineering best practices with the

Despite these intuitive frameworks, the vast majority of imaging modalities yield higher dimensional tensors than a standard RGB image. An echocardiogram for example, is 2D ultrasonographic video of the heart that can dimensionally be represented as $[channels, frames, x, y]$. CT and MRI-scans can be thought of as a z-stack of 2D images $[channels, z, x, y]$. These 'imaging' modalities thus yield tensors that are more similar to videos, where unstacking them as images may lead to the loss of spatial or temporal context: processing a video by analyzing each frame as a separate independent image for example, leads to the loss of temporal information between each video frame.[4,33,34] In a variety of tasks utilizing echocardiography, CT, and MRI scans, video based neural network algorithms have shown considerable improvements over their 2D counterparts.[2,4,35] Unlike the extensive libraries of pre-trained image based networks, support for video algorithms remains limited. Only three video neural network architectures with pretrained weights are currently available within Pytorch torchvision. Researchers interested in deploying newer architectures will likely need to perform pre-training steps on large publicly available video datasets (Kinetics, UFC) themselves.

While pre-training using large natural scenery datasets is an accepted strategy in developing clinical imaging ML systems, performance gains are not guaranteed.[36] While reports of performance improvements especially when working with smaller datasets are common with pre-training, the benefits taper off with larger training datasets.[2] As an alternative to pretraining on natural images or videos, Taleb *et. al*. propose a series of novel self-supervised pre-training techniques using large *unlabeled* medical imaging datasets, with the aim of assisting the development of 3D medical imaging based AI systems. Neural networks learn latent features representative of 3D medical images by training them to perform a set of 'proxy tasks'.[34] For example, by training a networks to 'reassemble' scrambled input scans, it is possible to learn a semantic representation of the data itself. Unlabeled imaging studies thus can serve as the groundwork for effective pretraining. This would be followed by fine-tuning on a smaller sample of high quality ground truth data towards a specific supervised learning tasks. While these steps help adapt existing neural network architectures for medical imaging, designing new architectures to specific tasks requires rare expertise. Advances in evolutionary search algorithms allow for machine learning methods to be used to discover multiple architectures tailored to a specific task, resulting in hyper-efficient and higher performance architectures than those constructed by humans.[37,38] These advances offer a unique opportunity towards modality specific architecture development.

Training deep learning algorithms rely on graphical processing units (GPUs) to perform the massively parallel matrix multiplication operations. The availability of cloud compute 'pay as you go' GPU resources, and consumer grade GPUs with high memory capacities have all helped reduce the barrier to entry for researchers interested in developing ML systems for medical imaging. Despite these advances, training complex modern network architectures on large video datasets requires multiple GPUs running for weeks.[30] Both TensorFlow and Pytorch have efficient and intuitive methods of training neural networks across multiple GPUs. Frameworks that extend their capabilities (Grid.ai, Horovod.ai) further allow for large scale distributed training with opportunistic resource allocation. These are especially useful when pre-training networks or optimizing hyperparameters on cluster or cloud compute environments. Powerful abstraction layers (Pytorch Lightning) now enable researchers to establish standards for structuring boilerplate ML code. Adopting a modular approach to designing machine learning systems will allow researchers to rapidly repurpose systems designed for various clinical imaging modalities, or extend the capabilities of these systems by integrating sub-components in novel ways.

**Time to event analyses and Uncertainty Quantification**

As medical AI systems shift from 'diagnostic' applications to more 'prognostic' applications, time-to-event predictions rather than simple binary predictions will find more relevance in the clinical setting. Time-to-event analyses are characterized by the ability to predict event probabilities as a *function* of time, whereas binary classifiers can only provide predictions for one predetermined duration. Unlike binary classifiers, time-to-event analyses account for censoring of data - where some individuals are either lost to follow up, or may not have experienced the event of interest the observation time-frame. Survival analyses are commonplace in clinical research, and are central to the development of evidence-based practice guidelines. Extending traditional survival models with image and video-based machine learning may provide powerful insight into the prognostic value of latent features embedded within histological sections or medical imaging scans. For example, integrating extensions of cox-proportional hazards loss functions into traditional neural network architectures enables the prediction of cancer outcomes from histopathology slides alone.[39,40] Critical for such translational efforts are the availability of robust and well validated deep learning extensions of the cox regression. Over the past years, a number of deep learning implementations of the cox model have been described. Kvamme *et. al.* propose a series of proportional and non-proportional extensions of the Cox model, with additional implementations of survival methods described in the past such as *DeepSurv* and *DeepHit*.[41] The models and associated loss functions are available in the pycox package, compatible with the pytorch ML framework. Similar implementations are available in TensorFlow and Caffe.

Humans tend to err on the side of caution when unsure about a task. This is somewhat mirrored by machine learning systems where the output is a 'class probability', or 'likelihood of being correct' between a scale of 0 to 1. Most medical imaging ML systems described in literature today, however, lack the implicit ability to say "I don't know" when provided input data that is out of distribution for the model. A classifier trained to predict pneumonia from CT scans for example, will provide an output (of either pneumonia or no pneumonia) even if the input image is that of a cat. In their paper on uncertainty quantification in deep learning, Sensoy *et. al.* addresses these issues with a series of loss functions that assign an 'uncertainty score' as a way to avoid erroneous but confident predictions.[42] Implementations of these methods are available under permissive licenses and compatible with commonly used ML frameworks. The incorporation of uncertainty quantification may help increase both the interpretability and the reliability of high stakes medical imaging ML systems, and reduce the likelihood of automation bias – a phenomenon where clinicians may over-rely on automation.[43]

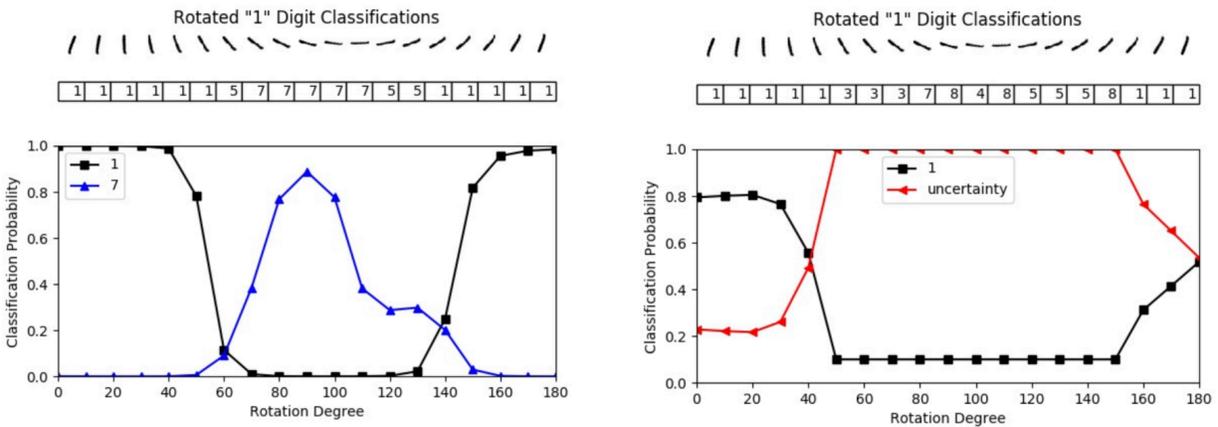

Figure 3. Machine learning models trained with standard methods can be extremely confident even when incorrect, as described by Sensoy et. al (reproduced with permission). As a digit is rotated 90 degrees, the system confidently assigns the label from '1' to '7'. On the other hand, with methods that account for classification uncertainty, the system assigns an 'uncertainty score' that can help alert clinicians to potentially erroneous predictions.

**Explainable AI and Risk of Harm**

Saliency maps and class activation maps remain the de-facto standard for explaining how ML algorithms make their predictions.[44,45] Adebayo *et al* recently describe how relying solely on the visual appearance of saliency maps can be misleading even if at first glance they appear contextually relevant. In a series of extensive tests they find that instead of deriving true meaning from model weights, many popular methods for generating post-hoc saliency maps are in face no different to 'edge detectors' (algorithms that simply map sharp transition areas between pixel intensities).[46] Furthermore even when these methods work, little can be deciphered beyond 'where' the ML algorithms are looking, with numerous examples where saliency maps look nearly identical for both correct and incorrect predictions.[47] These drawbacks are more pronounced when the difference between a 'diseased' state and a 'normal state' requires attention towards the same region of an image or video.[48] A more granular approach may involve serial occlusion tests, whereby performance is assessed on images after masking regions of the image that clinicians would otherwise use to make diagnoses or predictions.[49] This helps to confirm clinical intuition, or at times may show that ML systems are able to identify informative features from other areas of the scan that clinicians would traditionally not use.[6]

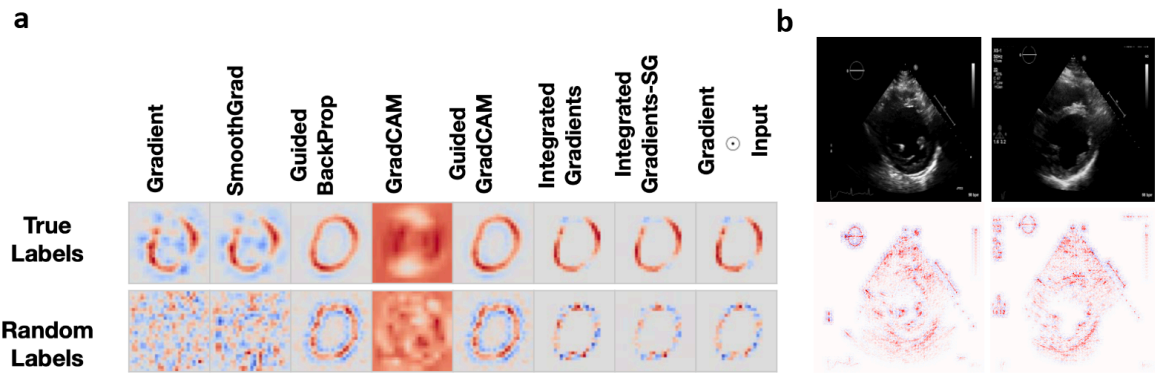

Figure 4a. Experiments conducted by Adebayo *et. al.* (reproduced with permission) with models trained on true labels from the MNIST dataset, and trained on random noise. Models trained on random noise still yield the circular shape of the digit zero when evaluated by the majority of visualization methods. These offer little in terms of true saliency maps, functioning more as class invariant edge detectors. b. Detection of echocardiographic view planes: both incorrect (left) and correct classifications (right) yield similar saliency maps.

The latent features a ML system can learn to some extent depends on architectural design choices. More importantly, ML systems will learn and perpetuate systemic inequities based on the training data and targets provided to it. As healthcare AI systems move towards future prediction of disease, greater care must be taken into accounting for the extensive disparities in access to healthcare and the outcomes across these groups. In a recent review, Chen *et al* give an in-depth review of potential sources of bias from problem selection to the post-deployment phase.[50] Here we focus on potential solutions early in the development of ML systems. As has been described previously, models may inadvertently learn to further perpetuate and discriminate against minorities and people of color.[50] While some have called for better methods to explain otherwise 'black box' predictions of modern ML systems, others advocate for more explainable models to begin with.[47] An intermediate approach involves training medical imaging neural networks using 'black box' models, in addition to incorporating inputs for structured data when training the overall AI system. This can be achieved by using secondary models or 'meta-learners', building fusion networks where tabular data is incorporated into one of the terminal fully connected layers of a convolutional neural network, or using autoencoders to create joint latent embeddings of images and structured data.[14,51,52] Trust in ML systems are critical for wider adoption, exploring how and why specific features or variables lead to predictions via a combination of saliency maps and model agnostic approaches of estimating feature importance.[53–55] Another newer approach is constraining a ML algorithm within the training logic, ensuring that optimization steps occur controlling for demographic variables of interest. From a technical standpoint, this would involve inserting an additional penalty loss in the training loop. Fairlearn for example, is a sci-kit learn toolkit to assess fairness in traditional ML models, and constrained optimizations based on the Fairlearn algorithms (FairTorch) have been developed that are a promising exploratory foray into incorporating bias adjustments *within* the training process.[56] Numerous open-source toolkits exist to help researchers determine the relative importance different variables and input streams (image predictions, and variables such as gender and race). These techniques may allow for the development of more equitable ML systems, and even uncover hidden biases where none are anticipated.[57]

**Conclusion**

While computational architectures and access to high quality data are key to building good models, integrating AI into medical practice requires proactive efforts to address bias, uncertainty, and explainability. As such, the skepticism surrounding medical imaging and AI is healthy, and for the most part, warranted. We hope that researchers will find this document useful, for both the overview of potential challenges that await them in terms of clinical deployment, but also the tacit guidance towards how some of these issues may be addressed at the stage of model development.